# The Nature of Surface States on Vicinal Cu (775): An STM and Photoemission Study

N. Zaki, K. Knox, R M. Osgood
Columbia University, New York, NY 10027, USA

P. D. Johnson,
Brookhaven National Lab, Brookhaven, NY, 11973, USA

J. Fujii, I. Vobornik and G. Panaccione
Istituto Officina dei Materiali - Consiglio Nazionale delle Ricerche (IOM-CNR)
Lab. TASC; S.S. 14, Km 163.5. I-34149 Trieste (Italy)

We report ARPES and a set of *in situ* STM measurements on a narrow-terrace-width vicinal Cu(111) crystal surface, Cu(775), whose vicinal cut lies close to the transition between terrace and step modulation. These measurements show sharp zone-folding (or Umklapp) features with a periodicity in $k_\parallel$, indicating that the predominant reference plane is that of Cu(775), i.e. that the surface is predominately step-modulated. Our measurements also show variation in Umklapp intensity with photon energy, which is consistent with prior ARPES experiments on other vicinal Cu(111) surfaces and in agreement with our designation of the state as being step modulated. The measurements also show a weak terrace-modulated state, which, based on several characteristics, we attribute to the presence of terrace widths larger than the ideal terrace width. By measuring the intensity ratio of the two distinct surface-state modulations from PE and the terrace-width distribution from STM, we derive a value for the terrace width, at which the surface-state switches between the two modulations.

# I. Introduction

Recently widespread interest has developed in understanding the detailed electronic structure of steps and step arrays on low-index vicinal-cut single-crystal surfaces [1]. On bare surfaces, regular step arrays can be formed with relative ease following sputtering and annealing of vicinal-cut metal crystals. In addition, the use of edge-defined growth on vicinal surfaces makes it possible to also form bimetallic one-dimensional systems [2-4]. These Angstrom-scale features can in turn be used to examine a wide variety of one-dimensional quantum phenomena [5]. In addition, typical step-terrace dimensions are sufficiently small, i.e. ~ 5-30 Å, to allow quantum-confined structures to be formed with binding energies that are much greater than the thermal energy at room temperature. In effect, step arrays and related Angstrom-scale surface patterning enable quantum-structured condensed matter systems to be fabricated *in situ*.

Because of this interest, a wide variety of studies have been devoted to understanding the fundamental electronic structure of stepped surfaces and how their electronic structure differs from that of low index flat surfaces. A particular focal point of this work has been electronic structure which derives from the well known Cu(111) surface states. Thus recently, extensive occupied photoemission studies on the surface state of a set of vicinal Cu(111) surfaces have been carried out[1, 6] and matched by KKR calculations [7]. In brief, ARPES measurements [8] have shown that two different classes of surface-state modulation exist: terrace-modulation and step-modulation. A terrace-modulated state is a (111) surface state confined by step risers and angle referenced (i.e. having its band minimum aligned) with respect to the terrace normal. A step-modulated state is a step-superlattice modulated state, due to the periodic potential of the steps, angle referenced to the macroscopic surface normal. ARPES measurements have showed that the surface-state class for a particular surface modulation depends strongly on vicinal angle. Thus surface-state modulation is terrace aligned if the vicinal angle is <5º and surface (or cut) aligned if the vicinal angle is >9º. The transition for this behavior was attributed in Refs. [1, 9] to the fact that at high vicinal angles the terrace-modulated state overlaps with bulk states (i.e. it becomes a resonance) and hence is less confined by surface features such as step potentials. The ARPES study also showed that Umklapp features occurred only for step-modulated electrons. The intensity of these features was found to vary with photon energy[10], a result attributed to matching of the final bulk electronic wavefunction with the evanescent tail of surface state wavefunction; this interpretation was verified by KKR calculations[7], which also explained the asymmetric appearance of the spectra with respect to the surface normal. Subsequent measurements [6] showed the importance of considering step-width disorder for interpreting spectra from large-terrace (i.e. <5°) vicinal surfaces.

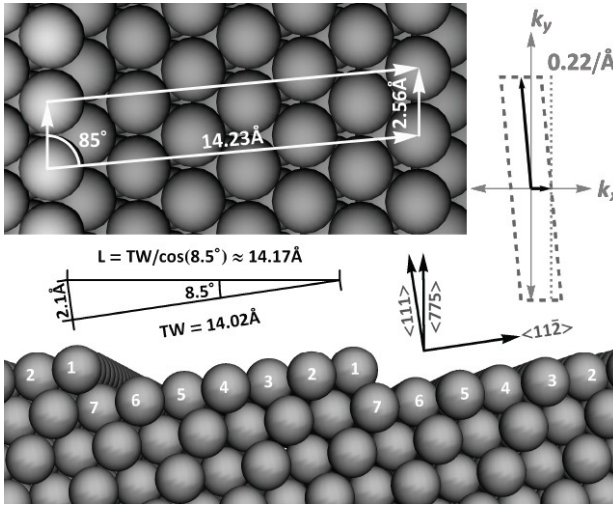

Fig. 1: Schematic diagram of Cu(775). The 7-row terrace is terminated by (111)-like edges; the steps are denoted as type-B steps, as is the case for all (n,n,n-2) surfaces; in contrast, (n,n,n+2) vicinal cuts are characterized by (100)-like edges and are denoted as type-A stepped surfaces. Due to the odd number of rows per step terrace, the unit cell is not rectangular, as in the case for even number of rows, but a parallelogram. The corresponding surface Brillouin zone is shown in the top right-hand corner, and its boundary in the direction normal to the step-edge is marked with a dotted line along with its wave-vector magnitude.

In this paper, we focus on using angle-resolved photoemission, in conjunction with *in situ* STM, to determine step-induced surface-state modulation for Cu(775). This vicinal surface is of particular importance since it is close to the transition point between the two different surface-state modulations and thus measurements on this substrate provide insight into the nature of the dominant modulation. These measurements show a variation in Umklapp intensity with photon energy indicating that the predominant reference plane is that of Cu(775), i.e. that the surface is predominately step-modulated. Our measurements also show a weak terrace-modulated state, which we attribute to the presence of terrace widths larger than the ideal terrace width. Finally we discuss the implications of these results to the current understanding of photoemission from surface states on vicinal surfaces.

## II. Experiment

The crystal surface examined in this paper is Cu(775), namely vicinal Cu(111) cut at 8.5°, which ideally consists of an array of type-B steps [11] with a step width of 6 1/3 atomic rows or approximately 14Å; see Fig. 1 for a diagram and relevant crystal orientation. Sample preparation, STM imaging of the stepped surface, and ARPES measurements were all performed *in situ* under UHV conditions with a base pressure ≤ $1 \times 10^{-10}$ Torr at the multi-instrument APE beamline [12] of the Elettra Synchrotron Facility in Trieste, Italy. This capability is essential for having both STM and photoemission carried out under well quantified conditions. For surface preparation, the electropolished sample was first subjected to repeated sputter and anneal cycles in order to remove bulk contaminants at the surface. These cycles consisted of sputtering at 2KeV at RT, annealing at 650ºC in $H_2$ at $10^{-6}$ Torr, and annealing at 650ºC in vacuum. A relatively uniform step array was obtained as shown in the STM image of Fig. 2 [4]. Second, the sample was re-cleaned daily using a mild sputter cycle of 1KeV and short annealing at a temperature of 450-475ºC. The sample was illuminated using the low-energy branch of the beamline with a photon energy range of 10-100eV and with p-linearly-polarized light. Room-temperature PE spectra were measured with a VG Scienta SES2002 electron-energy analyzer with an energy resolution of 10meV, an energy width lower than the expected thermal broadening, and with an angular resolution of 0.12°.

## III. Observations

Prior to and following photoemission measurements, *in situ* STM imaging was made of the surface. An example of this imaging is shown in Fig. 2 for one 50nm x 50nm region of the surface. Other regions were also examined *in situ*, as well as investigation of the surface preparation procedure *ex situ* in our laboratories at Columbia University. Figure 2 shows a regular step array displaying generally the dimensions expected for this surface, i.e. a terrace width of 14Å and a monolayer step riser; note that very large isolated (111) terraces were not observed. Because earlier studies have shown the importance of disorder in understanding stepped surface electronic structure, we carried out a statistical analysis of the images using an in-house imaging processing code that takes into account plane correction of a stepped surface, step-height/step-edge recognition, and average step-edge direction. This analysis enabled determination of the step width distribution shown graphically in the inset in Fig. 2. As has been observed previously[11, 13], due to the thermally activated step fluctuations and step-interactions, this surface has a distribution of terrace widths, which were characterized by a mean width of 14Å and a width range that extended to ~26Å.

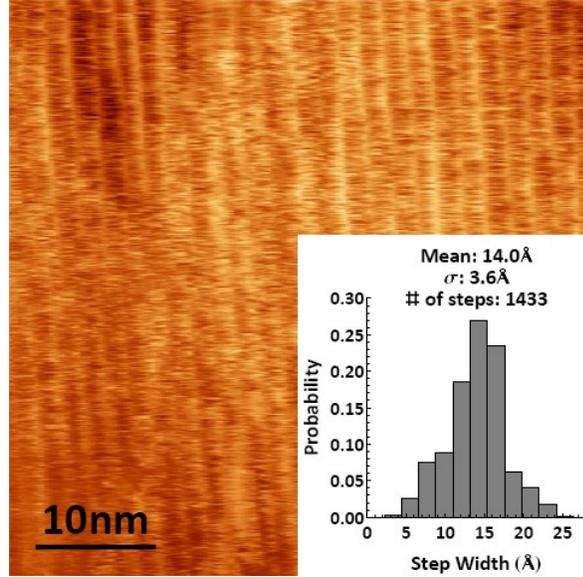

Fig. 2: (50nm x 50nm) RT STM image of Cu(775). The inset is a step-width distribution of Cu(775).

ARPES measurements at several photon energies spanning the range 15eV to 80eV were made on clean Cu(775). The measurements were performed for electrons emitted both perpendicular to and parallel to the step direction. The 1$^{st}$ surface Brillouin zone (SBZ), shown in Fig. 1, has boundaries of $\pm\pi/L = \pm 0.22$Å$^{-1}$ in the step perpendicular direction; repetitions or Umklapps of this zone exist in the step perpendicular direction with boundaries $\pm(\pi/L = 0.22$Å$^{-1}, 3\pi/L = 0.66$Å$^{-1})$, $\pm(3\pi/L = 0.66$Å$^{-1}, 5\pi/L = 1.10$Å$^{-1})$, … , $\pm((2n+1)\pi/L, (2n+3)\pi/L)$. Unless noted otherwise, the $(\pi/L = 0.22$Å$^{-1}, 3\pi/L = 0.66$Å$^{-1})$ Umklapp zone will be denoted as the 2$^{nd}$ zone and the $(3\pi/L = 0.66$Å$^{-1}, 5\pi/L = 1.10$ Å$^{-1})$ Umklapp zone will be denoted as the 3$^{rd}$ zone; note that these do not generally correspond to 2$^{nd}$ or 3$^{rd}$ SBZ's; note that the repeating zone scheme is used in the step perpendicular direction. In the case of step-perpendicular PE at low photon energy (i.e. < 46eV), the measurements clearly showed a parabolic dispersion curve whose minimum binding energy was centered at the 1$^{st}$-2$^{nd}$ zone boundary; see Fig. 3. This figure also includes the 2$^{nd}$ derivative of the ARPES data to accentuate the states. At higher photon energy, a replica of this feature is observed at a minimum binding energy centered at $3\pi/L$, corresponding to the 2$^{nd}$-3$^{rd}$ zone boundary. This feature is clearly an Umklapp replica of the surface state. For all photon energies the surface state and its Umklapp replicas were observed in only one half of the macroscopic surface plane; this half corresponds to the left side relative to [775], which includes the [111] direction as shown in Fig 1. The absence of a surface state and Umklapp bands observed symmetrically around $k_\parallel = 0$, i.e. bands do not appear for negative $k_\parallel$, is consistent with earlier measurements on other vicinal cuts [14] and the theoretical discussion by Eder, *et al*[7] who attribute this lack of symmetry to interference among the outgoing electrons. In particular, we observe half of the sp-derived surface state in the 1$^{st}$ zone, all of the first Umklapp in the 2$^{nd}$ zone, and half of the second Umklapp in the 3$^{rd}$ zone. To determine the binding energy, effective mass, and the linewidth of this state, photoemission data measured at 25eV is used since this procedure gives the best combination of momentum resolution and minimizes the overlap with an addition-al faint state, to be discussed below, relative to that at 15eV. The measured binding energy and linewidth is found to be 271 (±3) meV and 228 (±10) meV, respectively. These values agree well with our previous PE results on clean Cu(775)[15] and with the expected trend in linewidth, as reported in an earlier study of stepped surfaces[6]. The effective mass is measured to be ~0.57 $m_e$; this value is close to our previously measured value of ~0.47$m_e$[15].

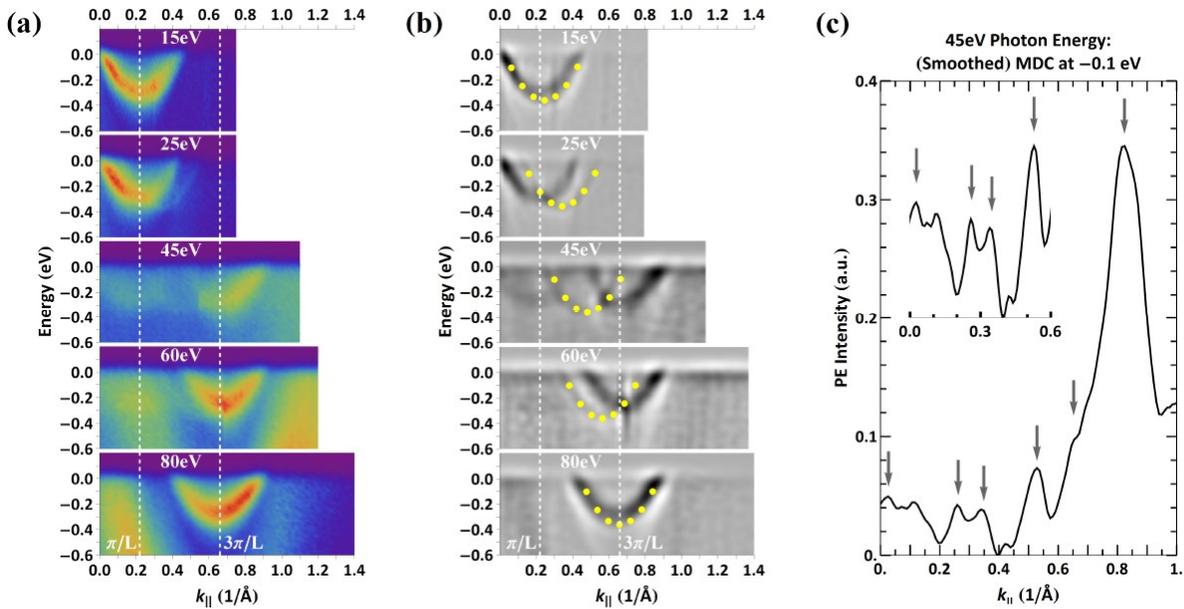

Fig. 3: (a) RT ARPES of clean Cu(775) in direction perpendicular to steps (i.e. -$k_x$). (b) 2$^{nd}$ derivative of ARPES data in (a). By varying the photon energy, the step-modulated surface state and its umklapp can be observed. As expected, the surface-state band minimum appears at the zone boundaries, as indicated by the white dashed vertical lines. Note that the faint state, indicated by yellow dots in (b), shifts with photon energy. This state is aligned with the step terrace normal, instead of the macroscopic surface normal, and hence is a terrace-modulated state. Apparently, for Cu(775), there are two competing modulations, step vs. terrace, as evidenced by the above ARPES measurements. (c) Example of a (smoothed) momentum distribution curve (MDC) taken at -0.1eV using ARPES data from 45eV photon energy; the marked peaks correspond to the step-modulated surface state, its Umklapp and the terrace-modulated state.

A striking effect, seen in Fig. 3, is the variation of the Umklapp intensity with the variation in photon energy. Thus, there is change in surface-state intensity in going from the combined 1st-2nd zone to the 2nd-3rd zone. For photon energies less than ~30eV, the relative surface-state intensity is highest in the 1st-2nd zone. For photon energies greater than ~40eV, the surface-state intensity is highest in the 2nd-3rd zone. Measurements of the variation in the PE intensity for photon energies between 30-40eV were not possible due to the weak photon emission intensity in this region. As mentioned above, the variation between the combined 1st-2nd zone and the combined 2nd-3rd surface zone has been discussed previously with respect to Au and Cu vicinal-cut surfaces. In Fig. 3b a 2nd derivative-plot of our data shows more clearly the surface state and its Umklapp. This filtered data reveals that at 45eV both curves cross the Fermi level before crossing with each other. This observation will be discussed later on in the analysis section.

The ARPES measurements in Fig. 3 also reveal a faint state that shifts in $k_{||}$ with photon energy; this feature is highlighted using yellow dots in Fig. 3b. Its effective mass is measured to be ~0.5$m_e$, that is, the same as the usual (111) surface-state value. As this state shifts in $k_{||}$, it is obscured at some photon energies by the stronger Umklapp feature as seen clearly in the data at 15eV and 80eV photon energies. Careful measurements show, however, that the binding energy is higher, ~360 meV, than the binding energy of the step-modulated state, and its linewidth is distinctly narrower, ≤ 120 meV. By examining the angle-space PE equivalent of Fig. 3, it is apparent that the band minimum of this state is aligned to the terrace normal. Since $k_{||} = \frac{\sqrt{2m_e E_{kin}}}{\hbar} \sin\theta$, the band minimum, located at 8.5°, will have $k_{||}$ change with photon energy, and hence the reason for the shift in $k_{||}$.

A careful consideration of the above data shows that characteristics of this weak state are those of a terrace-modulated state. First, the state's band minimum is aligned with the terrace normal, i.e. the [111] direction, instead of the macroscopic surface normal or the zone boundaries. Second, its larger binding energy of 360meV, compared to that of the step-modulated state, is that expected for a terrace-modulated state [6]. Third, this state does not show the characteristic Umklapp or bandfolding that is expected of step-modulation. The appearance of a weak terrace-modulated state raises the possibility that perhaps the surface preparation of Cu(775) left a mixed surface phase, that included minority regions of relatively large Cu(111) or macroscopic flat areas. However our STM imaging of the surface, including regions different than in Fig. 2 did not show any such regions; instead, as shown, for example, in the sample image of Fig. 2, the overall surface structure was relatively uniform across the sample. As mentioned earlier the finite step distribution seen in STM images is a consequence of the thermal activation at room temperature. The relatively uniform nature of our Cu(775) surface is reflected in the tight PE EDC linewidth; thus, we measure a linewidth of ~230meV, which is close to an expected linewidth of 180meV[6]. Note a measurement of the linewidth was complicated by the presence of the terrace-modulated state, which because of some residual overlap, gives the appearance of a larger linewidth.

The finite step distribution does play an important role in the presence and appearance of the faint step modulated state. The terrace-modulated state is characterized as a localized state on the terrace. It has been observed previously on Cu(665), a vicinal Cu(111) surface with mean terrace width of 25.2Å and a standard deviation derived terrace width range of 16.1Å to 34.3Å [6]. This prior work showed that in the presence of this distribution of terraces, these terraces, which individually support a specific quantum well state with flat energy dispersion, do exhibit an apparent free-electron like energy dispersion. Hence, this distribution changed the expected quantum-well energy dispersion to resemble a free-electron-like dispersion. As shown in Fig 2, the step-width distribution of Cu(775) includes terrace widths within the above cited range, albeit with lower probability compared to terrace widths close to the ideal of 14Å. To summarize, spatially on the surface, the predominate step-modulated state exists over an array of terrace-widths close to the ideal of 14Å. These step-modulated surface state supporting step-arrays are, however, interrupted by the larger width terraces in the step distribution, which support localized terrace-modulated states. Hence, the two surface-state modulations do not co-exist on the same step, but belong, respectively, to a step array of almost ideal step-width, and to a terrace of a larger step width.

The identification of the weak state as a terrace-modulate or localized state is further supported by the following observations. The weak state binding energy and linewidth are closer to that of Cu(665), mentioned above, than they are to flat Cu(111). In our case, the binding energy and linewidth are, respectively, ~360meV and ≤ 120meV; on Cu(665), they are ~350meV and 90meV; in the case of Cu(111), they are ~390meV and ≤ 60meV. As discussed further below, the ratio of the PE intensity of the second state to the combined PE intensity (predominate state plus weak state) is 0.22, suggesting that 22% of the sample area contributes to the second state. However, our STM studies on Cu(775) do not reveal large terraces that take up this percentage of the surface area. Hence, it is unlikely that source of the weak state signal is a few large terraces but rather a distribution of many terraces across the surface with widths large enough to hold a localized state.

In order to quantify better the presence of the weak terrace-modulated state, a measurement was made of the PE intensity ratio of the terrace-modulated to the step-modulated state. This ratio was measured at the band minima of the two states, taken at 25eV photon energy. Analysis at other photon energies was hindered by terrace-modulated-state/stepped-modulated state overlaps and by non-homogenous angle-space intensity mapping. We expect, however, that this ratio to be photon-energy independent, since photoemission intensity is approximately proportional to state population. The results of this measurement showed that the ratio was approximately 0.28 (0.02).

The presence of the terrace-modulated state in the vicinity of the angle for the surface-state-modulation transition, can lend direct experimental insight to the physics, which control this transition. If the terrace-modulated state PE signal is assumed to come from terrace-widths larger than the mean width, the above intensity ratio can be combined with a width-weighted terrace-width distribution to arrive at a value of the terrace width, at which the surface-state modulation switches. Using this approach, we calculate a value of 17.4Å (0.6) for the minimum terrace width. This value is close to the previously reported [8] transition terrace-width of 17Å. Furthermore, it is in agreement with the exclusively type-B vicinal Cu PE studies [6], in which 16.3Å average-terrace-widths were shown to be step-modulated while 25.2Å average-terrace-widths were shown to be terrace-modulated. This result should be of value in further theoretical studies of this transition point.

### III. Analysis and Discussion

In our discussion in this section, we first address the variation in surface-state intensity with photon energy. First and most important as stated above, the observation of photon-energy dependence for Umklapp features appearing at two different zone-edge momenta $k_\parallel = \pi/L$ and $3\pi/L$ is clear evidence that the (775) surface is step-modulated rather than terrace-modulated. This is an important result since it defines more precisely regions of vicinal-angle space, for which the transition described in Ref. [8] occurs.

The specific photon dependence measured in our experiments can be determined readily via use the free-electron approximation [16] for the final state to determine the perpendicular component of the final-state crystal momentum at a given photon energy. A diagram providing a PE transition-space representation that incorporates the free-electron approximation band is shown in Fig. 4. This diagram shows the expected [1, 6] initial step-modulated supperlatice state of ideal Cu(775) in white, and the free-electron final-state band in gray, along with direct transitions for a given photon energy, as denoted by vertical arrows. It is readily seen in the diagram that a transition from either the surface state or its Umklapp is possible at any photon energy. Hence, the reason for the observed intensity variation with photon energy is not due simply to the absence of an available final state at certain photon energies.

Ortega and his collaborators have provided insight into the origin of this alternating pattern of PE intensity with photon energy for the case of the stepped Au(322)[17] and later Cu(223)[10] by pointing out that it is a necessary consequence of photoemission resonances. These APRES resonances, which were first explained for the case of flat Cu(111) in Ref.[18], are the result of the periodicity in the evanescent portion of the full wavefunction in the direction normal to the surface. When inserted into Fermi's Golden Rule for the ARPES probability, this z-periodicity of the wavefunction gives rise to resonances for final-state free-electron wavefunctions of the same periodicity in $k_{perp}$ or its higher harmonics. Specifically, these conditions cause the final state k-vector to be $(2n+1)\,\pi/a_z\,k_{[111]}$, where $k_{[111]}$ is the unit k-space vector in the [111] direction. These resonances have been examined in the more recent theoretical study of photoemission on narrow-terrace stepped Cu(111) by Eder, *et al*[7] cited above, which used the screened KKR self-consistent method to investigate surface-state spectral resonances for both flat and stepped Cu(111), i.e. Cu(111), Cu(332) and Cu(221), respectively. For the sake of specificity, these authors focused on photon-energy resonances at the band minimum and showed that a series of resonances exist at the $k_\parallel$ discussed earlier by Ortega. For resonances on a *stepped* surface, the same basic physical effects lead to resonances as on a flat surface. However, due to the tilt of the surface with respect to the crystal plane, interference between the out-going electron with its initial state causes the final state that is parallel to the [111] direction, i.e. the local terrace normal instead of the surface normal, to have a finite amplitude. Thus the first resonance in photoelectron signal occurs at n=0 at low photon energies, e.g. ~0eV for Cu(775), which for this vicinal-cut sample corresponds to $k_\parallel = \pi/L$ or the 1$^{st}$-2$^{nd}$ zone boundary. The resonance for n=1 occurs at $k_\parallel=3\pi/L$, i.e. the 2$^{nd}$-3$^{rd}$ zone boundary at higher photon energy. The predicted photon energy for n=1, according to Ref. [7], is 69eV, a value which agrees with our experiment results. Note that the above discussion of intensity variation with $k_\parallel$ is based on the fact that our signal arises from a superlattice state and thus displays Umklapp replicas. Thus a second question regarding our results is why is such a superlattice state seen rather than the terrace-confined quantum-well state, which occurs at larger terrace widths.

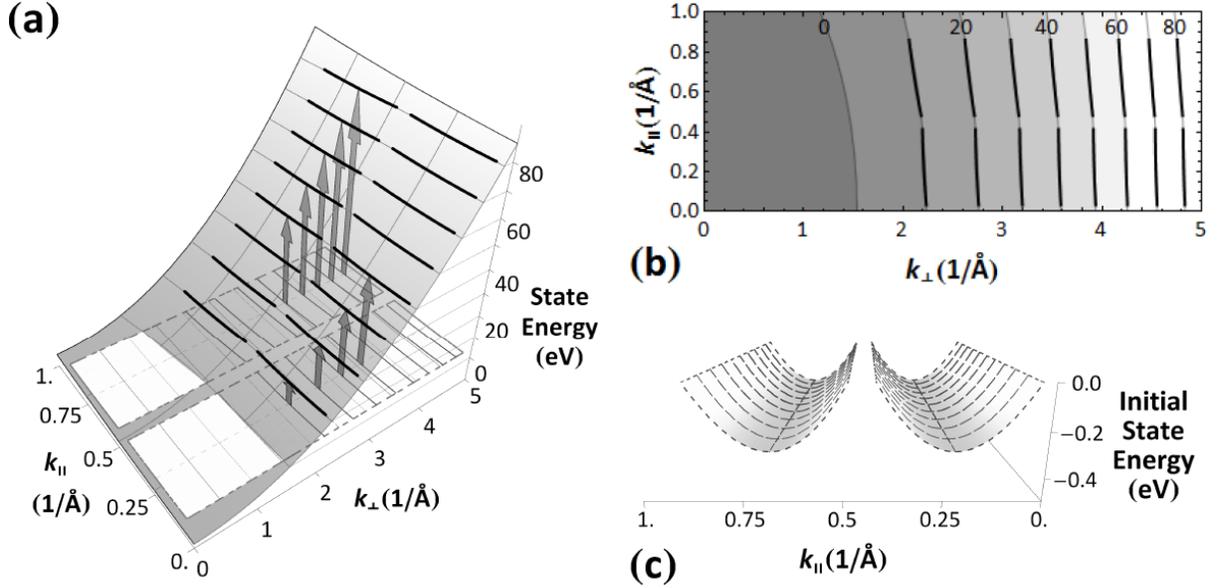

Fig. 4: (a) Schematic diagram of possible state transitions during angle-resolved photoemission from ideal Cu(775), i.e. perfect step-array of 14.0Å. All energies are referenced to the Fermi level. The initial state band, consisting of the surface state and its Umklapp, i.e. the step-modulated superlattice surface state, is denoted in white, centered at 0.22Å$^{-1}$ and 0.66Å$^{-1}$; their dispersion is shown clearly in (c). The final state band, a free-electron band approximation, is denoted in gray, and has a paraboloid shape; a contour plot of this band is shown in (b). Sample transitions for several photon energies are shown in (a) and (b) and some are explicitly marked by arrows in (a). The initial states are marked by thick gray lines on the initial state band in (a) and the final states are marked by thick black lines on the final state band in (a) and (b). Due to the low binding energy of the initial state, it and its corresponding final state appear to follow the lines of constant energy, as shown in (a) and (b).

One appealing explanation for the transition from step- to terrace-modulation that has been advanced is that a reduction of the projected band-gap with increasing vicinal angle causes the terrace-modulated surface state wavefunction to switch to step-modulated [9]. As a result it is important to confirm the presence of any band gap on our surface and to track its change with vicinal-cut angle. Accordingly tight-binding calculations [19] were used to calculate the projected gap with angle. In this calculation, a Slater-Koster tight-bonding interpolation scheme [20] was used to calculate the energy eigenvalues along any crystallographic direction using the eigenvalues calculated at high symmetry points from an APW (Augment-Plane-Wave) calculation; hence given a desired crystallographic direction, the k-space along that direction was adequately sampled to determine the band projection in that direction. To obtain the projection of the bulk bands, we follow the practice of other studies on vicinal surfaces (Au and Cu) of projecting the bands of the 1$^{st}$ Bulk BZ [1, 21, 22]. The result of band projections within the 1$^{st}$ Bulk BZ for three different (111) vicinal cuts, plus that for the flat surface, is displayed in Fig. 6. Clearly, the [111] band projected bandgap does not close or disappear for the (775) vicinal cut examined in this paper. Instead, its form remains approximately the same as for the flat (111) surface, though it does shift in $k_\parallel$ with increasing angle and, due to the tilt of the surface, is centered at the SBZ edge on the side of $k_\parallel$ which holds the [111] direction. For the case of small vicinal cuts (<5º), a terrace-modulated surface state (i.e. a surface state with a (111) reference plane), will fit within the projected bandgap [9]. However, for vicinal angles > 7.5º, bulk-states overlap completely with the projection of the terrace-modulated surface state [9]. This overlap will cause the terrace-modulated state to become a broad resonance and loose spectral weight [9] and cause it to be un-observable in an EDC of the surface. Further as shown in the figure and discussed earlier in [9] due to the fixed (111) orientation of the terrace, the step-modulated state remains in the projected bandgap for all the vicinal angles show in Fig. 6, thus giving a strong PE signal. While this explanation is very helpful in understanding the physics of the terrace/step modulation switch, it apparently does not fully explain why the step-modulated state is not observed for samples with smaller vicinal angles.

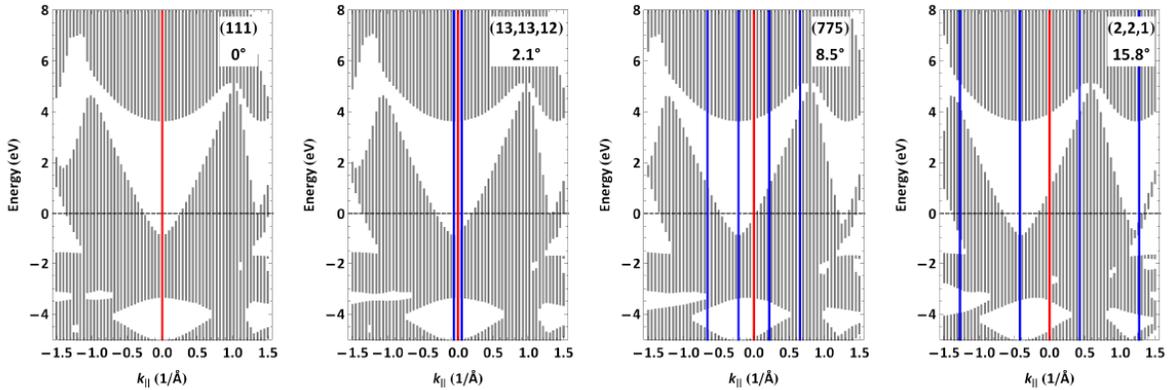

Fig. 5: Cu Bulk projection of 1$^{st}$ BZ for flat and vicinal Cu(111) surfaces. The sp surface state gap is at $k_\parallel$=0 for Cu(111) as expected. Note that the projection is not symmetrical with respect to $k_\parallel$; though the surface BZ of Cu(111) is 6-fold rotationally symmetric, the bulk BZ possesses only 3-fold rotational symmetry along the [111] direction which is reflected in the projection. For these vicinal surfaces, the sp gap shifts in the direction of $\overline{M}'$ (i.e. toward bulk symmetry point X), but does not close or lose noticeable volume with vicinal angle. The vertical red line denotes $k_\parallel$=0 which corresponds to the macroscopic surface normal. The blue vertical lines correspond to $\pm\pi/L$ and $\pm 3\pi/L$. Note that the sp gap minimum is aligned with the first surface BZ boundary, $-\pi/L$, as is the case with the step-modulated surface state. Also notice that there is no projected gap for the Umklapp surface state, centered at $-3\pi/L$. Furthermore, there is no projected gap for positive values of $k_\parallel$, i.e. $k_\parallel \geq 0$.

An alternate more recently advanced explanation for the switch from step to terrace modulation is based on measurements that suggest that this switch is a result of subtle changes in surface energetics[6]. This explanation rests on PE measurements of the opening of a bandgap in the superlattice state on Cu(443)[23] and concomitant rearrangement in the density of states below the Fermi level. For this surface, the authors of Ref. [23] observed a bandgap at the Fermi level in the step-modulated state, giving a calculated energy gain of ~11meV/unit cell compared to a gapless case. Cu(775) is very close to Cu(443) in terms of terrace width, 14Å vs. 16.2Å respectively, and hence would be expected to have a similar energy gain. Kronig-Penney modeling of the gap[23] showed that this energy gain persists as the terrace width increases until for terraces with width >26Å, the energy gain is replaced by an energy loss[23]. This loss of energy gain is due to the fact that the gap moves deep below the Fermi level. The energy gain of the superlattice or step modulated state in going from large to short terraces with increasing vicinal angle causes the step-modulated state to lie lower in energy and to dominate and thus for the PE signal from this state to be seen. Finally the authors of this work note that it is possible that energy conditions "might not be sufficient to explain the observed cross over" and they note that localization due to step disorder may also play a role in the transition from superlattice states[6]. Following the explanation based on the importance of a gap, we note again that Cu(775) is very close to Cu(443) in terms of terrace width and, in fact, a plot of energy gain versus average terrace size obtained in Ref. [23] shows that for the 14Å terraces of Cu(775) a similar energy gain of 10meV per unit cell is obtained. In the case of Cu(775), however, the gap lies above the Fermi level and thus is not possible to observe in photoemission. Thus our results are not inconsistent with an explanation based on a gap opening, although this explanation could not be confirmed with our experiments.

**Conclusion**

Our photon energy-dependent photoemission measurements on narrow-terrace-width vicinal Cu(775) crystal have enabled us to examine the transition region in vicinal-angle space for a change in the surface reference plane from predominately step-modulated to terrace-modulated. Our data also provide a clear example of the variation of the photoelectron signal intensity of the Umklapp features with photon energy, a variation, which supports the assignment of step modulation to the (775) surface. In addition, our measurements show evidence of a faint terrace-modulated state, further emphasizing the proximity of our vicinal cut to the transition angle be-tween terrace-modulation and step-modulation. This observation of a terrace-modulated state allowed us, in combination with STM measurements of the terrace-width distribution, to arrive at a value for the terrace-width at which the surface modulation switches. Our ARUPS identification of the surface-state class on Cu(775) is also consistent with recent discussions of the origin of the switch from step- to terrace-modulation in [6, 9]. Finally, we note the very recent publication of a PE/STM (ex-situ) study of vicinal noble metal step-width distributions and their interplay with the surface state [24].

## Acknowledgment

This research was supported by the Department of Energy Contract No. DE-FG 02-04-ER-46157. Work at Brookhaven National Laboratory was supported by the Department of Energy under Contract No. DE-AC02- 98CH10886. We thank Jerry Dadap, Mehmet Yilmaz, and Manolis Antonoyiannakis for several useful comments and suggestions. We thank Peter Lucak for writing some of the step-width analysis code.